\documentclass[%
 reprint,
showpacs,
 amsmath,amssymb,
 aps,
]{revtex4-1}

\usepackage{dcolumn}
\usepackage{bm}

\usepackage[dvipdfmx]{graphicx,color}

\usepackage{ulem} 

\newcommand{\average}[1]{\left\langle#1\right\rangle}

\def\<{\langle}
\def\>{\rangle}
\def\({\left(}
\def\){\right)}
\def\[{\left[}
\def\]{\right]}
\def\up{\uparrow}
\def\dn{\downarrow}

\def\e{\mathrm{e}}
\def\i{\mathrm{i}}
\def\d{\mathrm{d}}
\def\der{\partial}

\def\Im {{\mathrm{Im}} \,}
\def\+{\dagger}

\begin{document}


\title{Microscopic quantum interference in excitonic condensation of Ta$_2$NiSe$_5$}

\author{Koudai Sugimoto$^1$}
\author{Tatsuya Kaneko$^{2}$}
\author{Yukinori Ohta$^2$}%
\affiliation{%
$^1$Center for Frontier Science, Chiba University, Chiba 263-8522, Japan\\
$^2$Department of Physics, Chiba University, Chiba 263-8522, Japan
}%

\date{\today}

\begin{abstract}
The microscopic quantum interference associated with excitonic condensation in Ta$_2$NiSe$_5$ is studied in a BCS-type mean-field approximation.
We show that in ultrasonic attenuation the coherence peak appears just below the transition temperature $T_{\rm c}$, whereas in NMR spin-lattice relaxation the rate rapidly decreases below $T_{\rm c}$; these observations can offer a crucial experimental test for the validity of the excitonic condensation scenario in Ta$_2$NiSe$_5$.
We also show that excitonic condensation manifests itself in a jump of the heat capacity at $T_{\rm c}$ as well as in softening of the elastic shear constant, in accordance with the second-order phase transition observed in Ta$_2$NiSe$_5$.  
\end{abstract}


\maketitle



A prediction was made about half a century ago that in a semimetal or a narrow-gap semiconductor, electrons in a conduction band (CB) and holes in a valence band (VB) form pairs called excitons and the system spontaneously goes into a state of quantum condensation with macroscopic phase coherence~\cite{Mott1961PM, Knox1963SSP, Keldysh1965SPSS, Cloizeaux1965JPCS, Jerome1967PR, Halperin1968RMP, Kunes2015JPCM}.  
The state leads to the opening of a band gap in semimetals or to the flattening of the band edges in semiconductors, and is called the excitonic insulator state.
The phase of such states may generally be referred to as the excitonic phase.
A recent development in experimental techniques, in particular, angle-resolved photoemission 
spectroscopy (ARPES), enables one to observe the changes in the band structure due to possible 
excitonic condensation in some materials~\cite{Cercellier2007PRL, Monney2009PRB, Monney2012PRB, Wakisaka2009PRL, Wakisaka2012JSNM}.  
Thereby, the excitonic phases have attracted renewed attention in recent years~\cite{Kaneko2015PRB}.  

Here, we focus on a candidate material Ta$_2$NiSe$_5$ \cite{Wakisaka2009PRL}, which is a 
narrow-gap semiconductor undergoing a structural transition from an orthorhombic to monoclinic 
phase at $T_{\rm c} = 328$~K~\cite{DiSalvo1986JLCM, Canadell1987IC}.
The flat band was observed in the ARPES experiment~\cite{Wakisaka2009PRL, Wakisaka2012JSNM, Seki2014PRB}, which was interpreted to be due to excitonic condensation; i.e., a mean-field analysis of the proposed three-chain Hubbard model with electron-phonon coupling explains the simultaneous occurrence of excitonic condensation and structural transition \cite{Kaneko2013PRB}, and a variational-cluster-approximation calculation of the extended Falicov-Kimball model well reproduces the ARPES spectral weight observed experimentally \cite{Seki2014PRB}.
However, a ``smoking gun'' experiment that determines whether Ta$_2$NiSe$_5$ is really in the excitonic phase is still lacking.  

In this paper, we challenge this issue; i.e., we will argue that the presence/absence of the coherence peak caused by microscopic quantum interference in the state can provide a crucial experimental test for the validity of the excitonic condensation scenario in this material.
In the case of superconductivity, it is known that the coherence factors appearing in ultrasonic 
attenuation and nuclear-magnetic-resonance (NMR) relaxation rates played an essential role in confirming the validity of the BCS theory \cite{Bardeen1957PR,Hebel1959PR}.
We apply this concept to the case of excitonic phases, whose state can be written in terms of a BCS-type wavefunction as well.  
We thereby present how the microscopic quantum interference of the state can give rise to either a coherence peak or a rapid decrease in the temperature dependence of these rates, of which not much is known so far, except for a simple model calculation of the ultrasonic attenuation rate~\cite{Maki1971JLTP, Amritkar1978SSC}.  

In what follows, we will first introduce a three-chain Hubbard model with electron-phonon coupling 
that describes the low-energy electronic states of Ta$_2$NiSe$_5$, and present some generalization of the mean-field calculations of Kaneko \textit{et al.}~\cite{Kaneko2013PRB}, demonstrating an excitonic and structural phase transition~\cite{SM}.
We will then calculate the temperature dependence of the ultrasonic attenuation and NMR spin-lattice relaxation rates and demonstrate that the coherence peak appears in the ultrasonic attenuation rate due to constructive interference, while there occurs a rapid decrease in the NMR relaxation rate due to destructive interference, the behaviors of which are in contrast to those of BCS $s$-wave superconductivity.  
We will also carry out the calculations of thermodynamic quantities, such as heat capacity and elastic constant, and show that a jump is observed in the specific heat at the phase transition and that elastic softening relating to the structural phase transition is observed in the elastic shear constant.
Our theoretical predictions are in fair agreement with available experimental data obtained so far for Ta$_2$NiSe$_5$, which we hope will encourage further experimental studies to provide the proof that Ta$_2$NiSe$_5$ is in the excitonic phase.



\begin{figure}
\includegraphics[width=\linewidth]{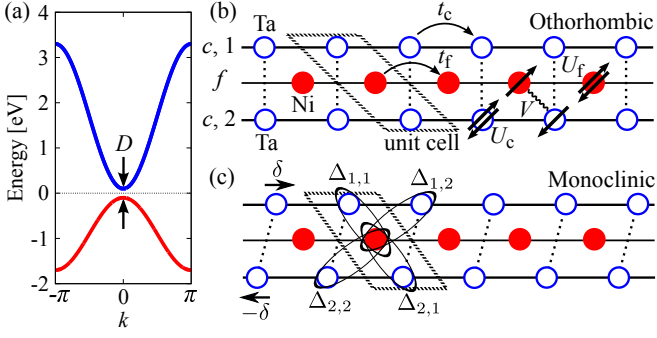}
\caption{
(a) Noninteracting band structure of our model with hopping parameters 
$t_{\rm c}=-0.8$ and $t_{\rm f}=0.4$~eV and band gap $D=0.2$~eV.  Also shown are 
the schematic representations of the lattice structures of (b) the orthorhombic and 
(c) monoclinic phases. 
We assume the on-site Coulomb repulsion $U_{\rm c}$ on $c$ and 
$U_{\rm f}$ on $f$ sites and intersite Coulomb repulsion $V$ between the $c$ and $f$ sites.  
We set $V=0.6$ eV and $U_{\rm c} = U_{\rm f} = 4V$. 
The electron-phonon coupling of the strength $\lambda=0.01$~eV is assumed unless otherwise stated. 
The order parameters are defined for the excitonic condensation as $\Delta_{\alpha, \beta}$ and for the structural distortion as $\delta$.  
}\label{fig:lattice}
\end{figure}

To be quantitative, let us introduce the three-chain Hubbard model with electron-phonon coupling used in Ref.~\cite{Kaneko2013PRB}, which consists of a doubly-degenerate CB of two Ta $5d$ chains and a nondegenerate VB of a Ni $3d$ chain, as is illustrated schematically in Fig.~\ref{fig:lattice}.
We define $c_{j, \alpha, \sigma}$ ($c_{j, \alpha, \sigma}^\+$) to be an annihilation (creation) operator of a CB electron at site $j$ with spin $\sigma$ on the chain $\alpha$ $(=1,2)$ and $f_{j, \sigma}$ ($f_{j, \sigma}^\+$) to be an annihilation (creation) operator of a VB electron at site $j$ with spin $\sigma$.
It was shown~\cite{Kaneko2013PRB,SM} that the BCS-type mean-field approximation to this model successfully describes the simultaneous occurrence of excitonic condensation and structural transition from the orthorhombic to monoclinic phase of Ta$_2$NiSe$_5$.  
The diagonalized mean-field Hamiltonian ${\cal H}^{\rm MF}$ reads 
\begin{equation}
 {\cal H}^{\rm{MF}}
	= \sum_{k, \sigma} \sum_{\epsilon={\rm c}, \pm}
		E_{k, \epsilon}
		\gamma^\dagger_{k, \epsilon, \sigma} \gamma_{k, \epsilon, \sigma}
 	+ N \epsilon_0,
\label{eq:Qa1hFNOH}
\end{equation}
where $E_{k, \epsilon, \sigma}$ is the quasiparticle energy, $\gamma_{k, \epsilon, \sigma}$ ($\gamma^\dagger_{k, \epsilon, \sigma}$) is an annihilation (creation) operator of the quasiparticle with band index $\epsilon$, wave number $k$, and spin $\sigma$, and $N \epsilon_0$ is a constant term in the mean-field Hamiltonian.
$N$ is the number of the unit cells in the system.  
The quasiparticle satisfies $c_{k, \mu, \sigma} = \sum_{\epsilon} \psi_{k, \sigma; \mu, \epsilon} \gamma_{k, \epsilon, \sigma}$, 
where $\psi_{k, \sigma; \mu, \epsilon}$ is the Bogoliubov transformation coefficient and $f_{k, \sigma} = c_{k, 3, \sigma}$.  
Details of our mean-field analysis are given in the Supplemental Material~\cite{SM}.  




First, let us discuss the ultrasonic attenuation rate.
Defining $A_{q, \alpha} = a_{q, \alpha} + a_{-q, \alpha}^\+$ using the phonon annihilation (creation) operator $a_{q, \alpha}$ ($a_{q, \alpha}^\+$) on the Ta chain $\alpha$, we write the Matsubara phonon Green's function as
\begin{equation}
 {\cal D}_{\alpha} (q, \tau) = - \average{{\rm T}_\tau A_{q, \alpha} (\tau) A_{-q, \alpha} (0)},
\end{equation}
where $A_{q, \alpha} (\tau) = \e^{- \omega_q \tau} a_{q, \alpha} + \e^{\omega_q \tau} a_{-q, \alpha}^\+$ is the Heisenberg representation of $A_{q, \alpha}$ at imaginary time $\tau$ and phonon wave number $q$ with a phonon dispersion $\omega_q$.
Using the Fourier coefficient ${\cal D}_{\alpha} (q, \i \omega_n) = \int^{\beta \hbar}_0 \d \tau \, {\cal D}_{\alpha} (q, \tau) \e^{\i \omega_n \tau}$, the phonon Dyson's equation is given as 
\begin{equation}
 {\cal D}_{\alpha} (q, \i \omega_n)
 	= {\cal D}^{(0)}_{\alpha} (q, \i \omega_n)
 		+ {\cal D}^{(0)}_{\alpha} (q, \i \omega_n) \Pi_{\alpha} (q, \i \omega_n) {\cal D}_{\alpha} (q, \i \omega_n),
\end{equation}
where $\Pi_{\alpha} (q, \i \omega_n)$ is the self-energy of the phonon Green's function.  
The ultrasonic attenuation rate is then given by the imaginary part of the retarded self-energy 
\cite{Hershfield1991PRB} as 
\begin{equation}
 \alpha_{q, \alpha}
 	= \frac{1}{\tau_{q, \alpha}}
 	= - 2 \, \Im \Pi_{\alpha}^{\rm R} (q, \omega_q + \i \eta),
\label{eq:9G8EYqqh}
\end{equation}
where $\tau_{q, \alpha}$ is a relaxation time of the phonon and $\eta$ is an infinitesimal value.

\begin{figure}
\includegraphics[width=0.95\linewidth]{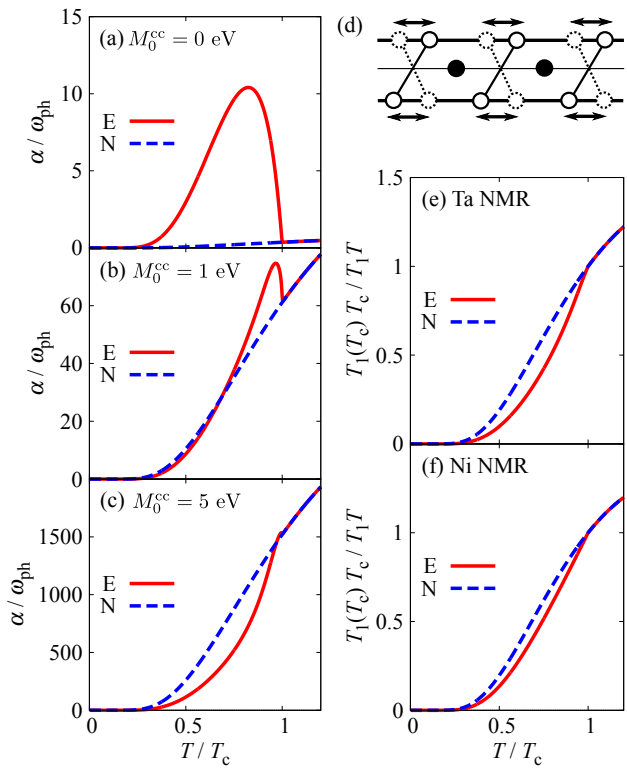}
\caption{
Calculated ultrasonic attenuation rate normalized by the ultrasonic frequency 
$\omega_{\rm ph}$.  We assume $M_0^{\rm cf}=1$~eV with 
(a) $M_0^{\rm cc} = 0$ eV, 
(b) $M_0^{\rm cc} = 1$ eV, and 
(c) $M_0^{\rm cc} = 5$ eV.  
(d) Schematic representation of the oscillation of the ultrasonic shear wave that propagates along the direction perpendicular to the chains.  Also shown are the calculated NMR relaxation rates at (e) Ta and (f) Ni sites.
The red solid line (E) is for the excitonic phase and the blue dashed line (N) is for the normal phase.  
}
\label{fig:response}
\end{figure}

We consider the lattice oscillation corresponding to the distortion in the structural phase transition, i.e., an ultrasonic shear wave for the transverse acoustic mode that propagates along the direction perpendicular to the chains [see Fig.~\ref{fig:response}(d)].  
The perturbation Hamiltonian of the phonons coupled with electrons is given by 
\begin{multline}
 {\cal H}'
	= \sum_{k, q, \sigma} \sum_{\alpha = 1}^2
	\biggl\{ M_{-q}^{\rm cc} A_{-q, \alpha}
		c^\+_{k_-, \alpha, \sigma} c_{k_+, \alpha, \sigma} \\
	- M_{-q}^{\rm cf}  A_{-q, \alpha}
		\( c^\+_{k_-, \alpha, \sigma} f_{k_+, \sigma}
		+ f^\+_{k_-, \alpha, \sigma} c_{k_+, \sigma} \) \biggr\} 
\label{eq:bpVNpQKq}
\end{multline}
with $k_{\pm} = k \pm \frac{q}{2}$.  
The first term represents the couping between the phonon and charge density of the CB electrons with a coupling constant $M_{-q}^{\rm cc}$, and the second term represents the hybridization between the conduction and valence bands by the phonon with a coupling constant $M_{-q}^{\rm cf} = \sqrt{\frac{ K \lambda \hbar}{M_{\rm c} \omega_{q}}}$, where $M_{\rm c}$ is the mass of a Ta ion, $K$ is the spring constant of the lattce harmonic oscillators, and $\lambda$ is the electron-phonon coupling strength~\cite{Kaneko2013PRB,SM}.
The coupling between the phonon and charge density of the VB electrons is ignored as being irrelevant to the present instability mode.
Equation~(\ref{eq:bpVNpQKq}) then reads
\begin{multline}
 {\cal H}'
	= \sum_{k, q, \sigma} \sum_{\alpha =1}^2 \sum_{\mu = 1}^3 W_{\alpha, \mu} (-q) A_{-q, \alpha} \\
		\times \( c^\+_{k_-, \alpha, \sigma} c_{k_+, \mu, \sigma} +  c^\+_{k_-, \mu, \sigma} c_{k_+, \alpha, \sigma} \)
\end{multline}
with $W_{\alpha, \mu} (q) = \( \frac{1}{2} M_{q}^{\rm cc} \delta_{\alpha, \mu} - M_{q}^{\rm cf} \delta_{\mu, 3} \)$.  

We adopt the second-order perturbation theory for the phonon self-energy.  
Since the ultrasonic wave number $q$ is small enough, we may assume $q \simeq 0$ in Eq.~(\ref{eq:9G8EYqqh}).  
We then obtain
\begin{align}
 &\alpha_{q=0, \alpha}
 	= 2 \pi \omega_{\rm ph} \sum_{k, \sigma} \sum_{\epsilon_1, \epsilon_2}
		\frac{\beta}{4 \cosh^2 \( \beta \( E_{k, \epsilon_1} - \mu \) / 2 \)}\,\times \notag \\
		& \delta \( E_{k, \epsilon_1} - E_{k, \epsilon_2} \)
		\sum_{\mu, \nu}
		W_{\alpha, \mu} W_{\alpha, \nu}
		\biggl\{ 2\Psi_{\alpha, \alpha; \epsilon_1} (k, \sigma) \Psi_{\mu, \nu; \epsilon_2} (k, \sigma) \notag\\
		&+\Psi_{\nu, \alpha; \epsilon_1} (k, \sigma) \Psi_{\mu, \alpha; \epsilon_2} (k, \sigma)
		+ \Psi_{\alpha, \mu; \epsilon_1} (k, \sigma) \Psi_{\alpha, \nu; \epsilon_2} (k, \sigma)
		\biggr\},
\end{align}
with $\Psi_{\nu, \kappa; \epsilon} (k, \sigma) = \psi_{k, \sigma; \nu, \epsilon} \psi_{k, \sigma; \kappa, \epsilon}^*$ 
and the ultrasonic frequency $\omega_{\rm ph}$.  

The calculated results for the temperature dependence of the ultrasonic attenuation rate are shown in Figs.~\ref{fig:response}(a)-~\ref{fig:response}(c).  We find the following:
In the normal state (obtained with vanishing order parameters), thermally excited electrons are scattered by phonons via the coupling with charge density, resulting in the behavior $\alpha\propto(M_0^{\rm cc})^2$.
The $M_0^{\rm cf}$ term does not contribute here.  
In the excitonic phase, a large coherence peak appears due to the phonon-induced $c$-$f$ hybridization ($M_0^{\rm cf}$), which is however overwhelmed by the charge-density term ($M_0^{\rm cc}$) at $M_0^{\rm cc}\gg M_0^{\rm cf}$, where the increase in the band gap suppresses the thermal excitation of electrons, resulting in a rapid decrease in the rate $\alpha$.  
However, in the ultrasonic attenuation experiment using the transverse sound mode, the coupling between the phonon and charge density of electrons does not contribute to the rate, and therefore we have the situation shown in Fig.~\ref{fig:response}(a).
The experimental observation of the coherence peak in the ultrasonic attenuation rate should thus be realizable.


Next, let us discuss the NMR spin-lattice relaxation rate, which may be written 
\cite{Moriya1963JPSJ} as 
\begin{equation}
  \frac{1}{T_{1,\mu}}
	\propto
	- \frac{k_{\rm B} T}{\hbar \omega_\mu}  \sum_q
	\Im \chi_{+-, \mu}^{\rm R} (q, \omega_\mu )
\label{eq:QReFp3KJ}
\end{equation}
using the transverse dynamical spin susceptibility
\begin{equation}
 \chi_{+-, \mu}^{\rm R} (q, \omega_\mu )
	= - \i \int^\infty_{-\infty} \d t \, \e^{\i \omega_\mu t} \average{ \[ S_{q, \mu}^+ (t), S_{-q, \mu}^- (0) \]} \theta (t), 
\end{equation}
where we define $S_{q, \mu}^+ = \sum_{k} c_{k_-, \mu, \up}^\+ c_{k_+, \mu, \dn}$ and 
$S_{q, \mu}^- = \sum_{k} c_{k_-, \mu, \dn}^\+ c_{k_+, \mu, \up}$, and $\omega_{\mu}$ is 
a resonant frequency of nuclear spins ($\mu=1,2$ for Ta and $\mu=3$ for Ni).  
Using the mean-field approximation and assuming a small $\omega_{\mu}$ value 
compared to typical energy scales of the system, we may rewrite Eq.~(\ref{eq:QReFp3KJ}) as 
\begin{multline}
 \frac{1}{T_{1,\mu}}
	\propto \pi
		\sum_{k, q} \sum_{\epsilon_1, \epsilon_2}
		\Psi_{\mu, \mu; \epsilon_1} (k_-, \up)
		\Psi_{\mu, \mu; \epsilon_2} (k_+, \dn) \\
	\times \frac{1}{4 \cosh^2 \( \beta \( E_{k_{-}, \epsilon_1} - \mu \) / 2 \)}
		\delta \( E_{k_-, \epsilon_1} - E_{k_+, \epsilon_2} \).  
\end{multline}

The calculated results for the temperature dependence of the NMR relaxation rate are shown in Figs.~\ref{fig:response}(e) and \ref{fig:response}(f) for Ta and Ni nuclear spins, respectively.
We find that, in contrast to the typical $s$-wave superconducting phase~\cite{Hebel1959PR}, 
there appear no characteristic peaks in the rate of the excitonic phase but the rate simply drops just below $T_{\rm c}$.
Thus, the behavior of the NMR relaxation rate in the excitonic phase is similar to that of an ultrasonic attenuation rate in the $s$-wave superconducting phase.  We point out that a recent NMR experiment on Ta$_2$NiSe$_5$~\cite{Lee2015} suggests the behavior of the rate consistent with our theoretical prediction.


Here, we briefly mention the nuclear-quadrupole-resonance (NQR) relaxation rate.  
We first note that the quadrupole interaction in NQR is the BCS case I interaction (without spin-flip processes)~\cite{Hammond1960PR} while the spin-lattice relaxation in NMR is the BCS case II interaction (with spin-flip processes)~\cite{Bardeen1957PR, Tinkham1975}.
Because the electron-phonon interaction in the ultrasonic attenuation is the Case I interaction as well, we may expect that the NQR relaxation rate should behave similarly with the ultrasonic attenuation rate where the coherence peak rapidly grows below $T_{\rm c}$, as we have shown 
above.
However, because the nuclear-quadrupole interaction comes not only from the on-site anisotropic charge distribution (which does not cause the coherence peak) but also from the intersite quadrupole interaction and the latter is not easy to evaluate, here we only suggest the possibility that the coherence peak can appear in the NQR relaxation rate, just as in the ultrasonic attenuation rate shown in Figs.~\ref{fig:response}(a)-~\ref{fig:response}(c).
Experimental studies are desired.  

\begin{figure}
\includegraphics[width = \linewidth]{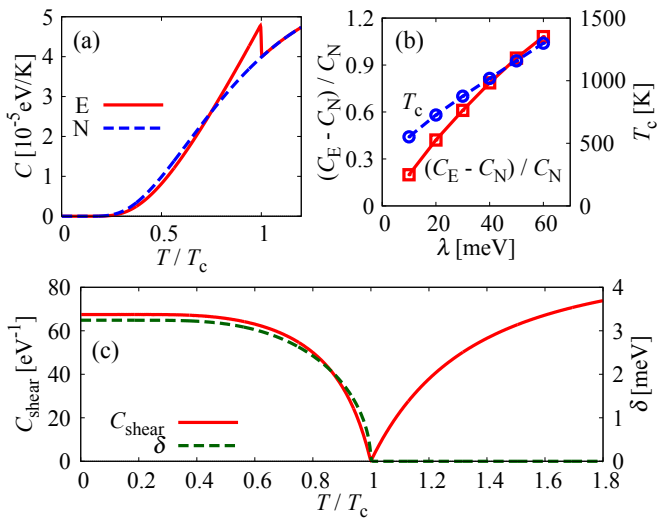}
\caption{
(a) Calculated temperature dependence of the heat capacity per unit cell, where the red solid line is for the excitonic phase and blue dashed line is for the normal phase. 
(b) Calculated $\lambda$ dependence of the jump in the heat capacity (red squares) and $T_{\rm c}$ (blue circles).  
(c) Calculated temperature dependence of the elastic shear constant per unit cell (red solid line) and that of the lattice distortion $\delta$ (green dashed line).
}
\label{fig:c}
\end{figure}


Finally, let us demonstrate that the excitonic condensation manifests itself in some thermodynamic quantities, such as heat capacity and elastic constant, which will provide additional experimental support for the validity of the excitonic condensation scenario in Ta$_2$NiSe$_5$.
Let us first discuss the heat capacity, which may be calculated from the mean-field free energy \cite{SM} as 
\begin{equation}
 C = - T \frac{\der^2 F}{\der T^2}
	= \sum_{k, \epsilon, \sigma}
		\( E_{k, \epsilon} - \mu \) \frac{\der f(E_{k, \epsilon})}{\der T},
\end{equation}
where $f(E)$ is the Fermi distribution function.  
The calculated result is shown in Fig.~\ref{fig:c}(a), where we find that the jump at $T_{\rm c}$ associated with the second-order phase transition is clearly visible, satisfying the entropy balance.  
The jump is given by $\( C_{\rm E} - C_{\rm N} \) / C_{\rm N} \simeq 0.20$ for the parameter values appropriate for Ta$_2$NiSe$_5$, where $C_{\rm E}$ and $C_{\rm N}$ are the heat capacities in the excitonic and normal phases, respectively, at $T_{\rm c}$.  This value is much smaller than the value 1.43 (a universal constant) in the BCS superconductivity and depends strongly on the model parameters used; its $\lambda$ dependence, e.g., is shown in Fig.~\ref{fig:c}(b).
Such a difference in the magnitude of the jump comes mainly from the difference in the normal phase:
It is a band insulator in the present excitonic condensation while it is a metal in superconductivity.
We also note in Fig.~\ref{fig:c}(b) that the jump in the heat capacity and the value of $T_{\rm c}$ increase monotonically as $\lambda$ increases, indicating that the larger values of the order parameters in the excitonic phase lead to the larger jump in the heat capacity.
We point out that a recent specific heat measurement on Ta$_2$NiSe$_5$~\cite{Lu2014} reveals a behavior consistent with our theoretical prediction.  


The elastic shear constant may also be calculated from the mean-field free energy~\cite{SM} as 
\begin{equation}
C_{\rm shear} = \frac{\partial^2 F}{\partial \delta^2}, 
\end{equation}
where we assume the lattice distortion of the transverse acoustic phonon mode in the long wavelength limit, corresponding to the observed structural phase transition [see Fig.~\ref{fig:response}(d)].
The calculated result is shown in Fig.~\ref{fig:c}(c), where we actually find the elastic softening at $T_{\rm c}$, leading to the structural phase transition.
We observe a Curie-Weiss--like  behavior $1/C_{\rm shear}=1/C_{\rm shear}^\infty+A/(T-T_{\rm c})$ at $T>T_{\rm c}$ with $1/C_{\rm shear}^\infty=0.094$~eV and $A=0.546$~eV~K.
A recent experimental observation of the diffuse x-ray scattering \cite{Sawa2014} suggests the presence of the soft phonon mode, which is consistent with our theoretical prediction.  The calculated lattice distortion $\delta$ at $T<T_{\rm c}$ is also consistent with the observed temperature dependence of the monoclinic angle of the lattice \cite{DiSalvo1986JLCM}.  



In summary, we have studied the microscopic quantum interference associated 
with excitonic condensation in Ta$_2$NiSe$_5$ using the BCS-type mean-field 
approximation for the three-chain Hubbard model with electron-phonon coupling.  
We have calculated the temperature dependence of the ultrasonic attenuation and 
NMR relaxation rates and have shown that the coherence peak can appear in the 
ultrasonic attenuation rate just below $T_{\rm c}$.  
In the NMR relaxation rate, on the other hand, no characteristic peak appears in 
$1/T_1$ but it simply drops just below $T_{\rm c}$, in agreement with recent 
NMR data for Ta$_2$NiSe$_5$ \cite{Lee2015}.  
The direct observation of the coherence peak in the ultrasonic attenuation rate 
will then provide a crucial experimental test for the presence of the excitonic 
phase in Ta$_2$NiSe$_5$.  
We have also demonstrated that the heat capacity exhibits a relatively small 
jump at $T_{\rm c}$ and the elastic shear constant indicates a softening when 
the temperature approaches $T_{\rm c}$, both of which are consistent with recent 
experimental observations for Ta$_2$NiSe$_5$ \cite{Lu2014,Sawa2014}.
We therefore hope that our theoretical predictions made here will encourage further 
experimental studies to provide proof that the excitonic condensation actually 
occurs in Ta$_2$NiSe$_5$.  

We thank M. Itoh, Y. Kobayashi, Y. F. Lu, K. Matsubayashi, T. Mizokawa, H. Sawa, 
and H. Takagi for discussing experimental aspects of Ta$_2$NiSe$_5$ and H. Fehske, H. Fukuyama, M. Ogata, and T. Toriyama for theoretical aspects.
This work was supported in part by Grants-in-Aid for Scientific Research from 
JSPS (No.~26400349 and No. 15H06093) of Japan.
T.K. acknowledges support from a JSPS Research Fellowship for Young Scientists.  

\nocite{*}



\end{document}


\title{Supplemental Material for 
\\``Microscopic quantum interference in excitonic condensation of Ta$_2$NiSe$_5$''}

\author{Koudai Sugimoto$^1$}
\author{Tatsuya Kaneko$^{2}$}
\author{Yukinori Ohta$^2$}%
\affiliation{%
$^1$Center for Frontier Science, Chiba University, Chiba 263-8522, Japan\\
$^2$Department of Physics, Chiba University, Chiba 263-8522, Japan
}%

\maketitle

\renewcommand{\theequation}{S.\arabic{equation}}
\renewcommand{\thefigure}{S\arabic{figure}}

In this Supplemental Material, we present the mean-field analysis of the three-chain Hubbard model 
with electron-phonon coupling, whereby we make a slight generalization of the analysis given in 
Ref.~\cite{Kaneko2013PRB}.  
The Hamiltonian of our model is written as $H = H_{0} + H_{\rm e-e} + H_{\rm lat}$, where 
$H_{0}$ is a kinetic term, $H_{\rm e-e}$ is an electron-electron interaction term, and 
$H_{\rm lat}$ is a lattice-distortion term relating to the structural phase transition in Ta$_2$NiSe$_5$.  
The number operators of the conduction band (CB) and valence band (VB) electrons are defined as 
$\hat{n}^{\rm c}_{j, \alpha, \sigma} = c^\+_{j, \alpha, \sigma} c_{j, \alpha, \sigma}$ and 
$\hat{n}^{\rm f}_{j, \sigma} = f^\+_{j, \sigma} f_{j, \sigma}$, respectively.
The kinetic term is written as
\begin{equation}
 H_0
	= t_{\rm c} \sum_{j, \alpha, \sigma} \sum_{\delta = \pm1} c_{j + \delta, \alpha, \sigma}^\+ c_{j, \alpha, \sigma}
		+ \varepsilon_{\rm c} \sum_{j, \alpha, \sigma} \hat{n}^{\rm c}_{j, \alpha, \sigma}
		+ t_{\rm f} \sum_{j, \sigma} \sum_{\delta = \pm1}  f_{j + \delta, \sigma}^\+ f_{j, \sigma}
		+ \varepsilon_{\rm f} \sum_{j, \sigma} \hat{n}^{\rm f}_{j, \sigma},
\end{equation}
where $t_{\rm c(f)}$ and  $\varepsilon_{\rm c(f)}$ are the hopping integrals and on-site energies of the CB (VB) electrons, respectively.
The electron-electron interaction term is expressed as $H_{\rm e-e} = H_{U_{\rm c}} + H_{U_{\rm f}} + H_{V}$ with 
\begin{equation}
 H_{U_{\rm c}} = U_{\rm c} \sum_{j, \alpha} \hat{n}_{j, \alpha, \up}^{\rm c} \hat{n}_{j, \alpha, \dn}^{\rm c}, \quad
 H_{U_{\rm f}} =  U_{\rm f} \sum_{j} \hat{n}_{j, \up}^{\rm f} \hat{n}_{j, \dn}^{\rm f},
\end{equation}
and
\begin{equation}
 H_V
	= V \sum_{j, \sigma, \sigma', \alpha} \hat{n}_{j, \alpha, \sigma}^{\rm c} \hat{n}_{j, \sigma'}^{\rm f}
		+ V \sum_{j, \sigma, \sigma'} \( \hat{n}_{j + 1, 1, \sigma}^{\rm c} + \hat{n}_{j - 1, 2, \sigma}^{\rm c} \)
			\hat{n}_{j, \sigma'}^{\rm f}, 
\end{equation}
where $H_{U_{\rm c (f)}}$ is the on-site Coulomb interaction $U_{\rm c(f)}$ of the CB (VB) electrons, 
and $H_{V}$ is the intersite interaction $V$ between the CB and the VB electrons.
The lattice-distortion term consists of a term of harmonic oscillators for the Ta atoms and 
a hybridization term between CB and VB electrons caused by the lattice distortion, and is written as
\begin{equation}
 H_{\rm lat}
	= \sum_{j, \alpha}
		\left\{ \frac{K}{2} X_{j, \alpha}^2 
		- \gamma_\alpha X_{j, \alpha}
		\sum_{\sigma} \( c_{j, \alpha, \sigma}^\+ f_{j, \sigma} + f_{j, \sigma}^\+ c_{j, \alpha, \sigma} \) \right\},
\label{eq:oGXUgl3k}
\end{equation}
where $X_{j, \alpha}$ is the displacement of a Ta atom at site $j$ in the $\alpha$-th chain measured 
from its equilibrium position, and $K$ is a spring constant of the harmonic oscillators.  
$\gamma_\alpha$ is the strength of hybridization between the CB and VB electrons caused by the 
lattice distortion.  We assume $\gamma_1 = \gamma$ and $\gamma_2 = -\gamma$.

Assuming the uniform electron distribution, the numbers of electrons per Ta and Ni atom may 
be written, respectively, as
\begin{equation}
 \average{c_{j, \alpha, \sigma}^\+ c_{j, \alpha, \sigma}} = n_{\rm c}, \quad
 \average{f_{j, \sigma}^\+ f_{j, \sigma}} = 1 - 2n_{\rm c}.  
\end{equation}
The excitonic order parameters (or excitonic gap functions) may be defined by
\begin{align}
 \Delta_{1, 1} &= V \average{c_{j, 1, \sigma}^\+ f_{j, \sigma}}, \quad
 \Delta_{1, 2} = V \average{c_{j + 1, 1, \sigma}^\+ f_{j, \sigma}},
\\
 \Delta_{2, 1} &= V \average{c_{j, 2, \sigma}^\+ f_{j, \sigma}}, \quad
 \Delta_{2, 2} = V \average{c_{j - 1, 2, \sigma}^\+ f_{j, \sigma}},
\end{align}
and the order parameter of the uniform lattice distortion may be defined by
\begin{equation}
 \delta = \gamma_\alpha X_{j, \alpha} = \gamma X.
 \label{eq:60EY9EJM}
\end{equation}
These order parameters are determined to minimize the free energy.  
Note that the definition of the excitonic gap functions is slightly generalized from the previous 
study \cite{Kaneko2013PRB} but the calculated results do not change significantly.  
We make use of the parameter $\lambda = \frac{\gamma^2}{2 K}$ (instead of $\gamma$) 
for the strength of the electron-lattice coupling below as well as in the main text.    
Using the order parameters thus defined, the electron-electron interaction term of the 
Hamiltonian may be written in the mean-field approximation as follows: 
For the Ta chain it reads
\begin{equation}
 H_{U_{\rm c}}^{\rm MF}
	= U_{\rm c} n_{\rm c} \sum_{j, \alpha, \sigma} \hat{n}_{j, \alpha, \sigma}^{\rm c}
		- 2 U_{\rm c} N n_{\rm c}^2,
\end{equation}
for the Ni chain it reads
\begin{equation}
H_{U_{\rm f}}^{\rm MF}
 	= U_{\rm f} \( 1 - 2 n_{\rm c} \) \sum_{j, \sigma} \hat{n}_{j, \sigma}^{\rm f}
		- U_{\rm f} N \( 1 - 2 n_{\rm c} \)^2,
\end{equation}
and for the intersite term it reads
\begin{align}
 H_V^{\rm MF}
	=& 8 V n_{\rm c} \sum_{j, \sigma} \hat{n}_{j, \sigma}^{\rm f}
		+ 4 V \( 1 - 2 n_{\rm c} \) \sum_{j, \sigma, \alpha} \hat{n}_{j, \alpha, \sigma}^{\rm c}
		- 16 V N n_{\rm c} \( 1 - 2 n_{\rm c} \)
		+ \frac{2N}{V} \sum_{\alpha, \mu} \left| \Delta_{\alpha, \mu} \right|^2 \notag \\
		&- \sum_{j, \sigma} \Bigl\{
			\Delta_{1, 1} f_{j, \sigma}^\+ c_{j, 1, \sigma}
			+ \Delta_{1, 2} f_{j, \sigma}^\+ c_{j + 1, 1, \sigma}
			+ \Delta_{1, 1}^* c_{j, 1, \sigma}^\+ f_{j, \sigma}
			+ \Delta_{1, 2}^* c_{j + 1, 1, \sigma}^\+ f_{j, \sigma}
			\Bigr\} \notag \\
		&- \sum_{j, \sigma} \Bigl\{
			\Delta_{2, 1} f_{j, \sigma}^\+ c_{j, 2, \sigma}
			+ \Delta_{2, 2} f_{j, \sigma}^\+ c_{j - 1, 2, \sigma}
			+ \Delta_{2, 1}^* c_{j, 2, \sigma}^\+ f_{j, \sigma}
			+ \Delta_{2, 2}^* c_{j - 1, 2, \sigma}^\+ f_{j, \sigma}
			\Bigr\},
\end{align}
where $N$ is the number of the unit cells.

Defining the Fourier transforms of the annihilation and creation operators as 
\begin{equation}
 c_{j, \mu, \sigma} = \frac{1}{\sqrt{N}} \sum_{k} \e^{\i k j a} c_{k, \mu, \sigma}, \quad
 c_{j, \mu, \sigma}^\+ = \frac{1}{\sqrt{N}} \sum_{k} \e^{- \i k j a} c_{k, \mu, \sigma}^\+,
\end{equation}
we may write the mean-field Hamiltonian in $k$ space as 
\begin{equation}
 H^{\rm MF}
 	=\sum_{k, \alpha, \sigma}
		\tilde{\varepsilon}_{\rm c} (k) c_{k, \alpha, \sigma}^\+ c_{k, \alpha, \sigma}
		+ \sum_{k, \sigma} \tilde{\varepsilon}_{\rm f} (k)
			f_{k, \sigma}^\+ f_{k, \sigma}
		- \sum_{k, \sigma, \alpha} \left\{
		\tilde{\Delta}_{\alpha} (k) f_{k, \sigma}^\+ c_{k, \alpha, \sigma} 
		+ \tilde{\Delta}_{\alpha}^* (k) c_{k, \alpha, \sigma}^\+ f_{k, \sigma}
		\right\}
		+ N \epsilon_0,
\label{eq:mwjFOxAS}
\end{equation}
with 
\begin{subequations}
\begin{gather}
 \tilde{\varepsilon}_{\rm c} (k) = \varepsilon_{\rm c} (k) + U_{\rm c} n_{\rm c} +  4 V \( 1 - 2n_{\rm c} \), \qquad
 \tilde{\varepsilon}_{\rm f} (k) =  \varepsilon_{\rm f} (k) + U_{\rm f} \( 1 - 2n_{\rm c} \) + 8 V n_{\rm c}, \\
 \tilde{\Delta}_{1} (k) = \Delta_{1, 1} + \Delta_{1, 2} \e^{\i k a} + \delta, \qquad
 \tilde{\Delta}_{2} (k) = \Delta_{2, 1} + \Delta_{2, 2} \e^{- \i k a} + \delta, \\
 \varepsilon_{\rm c} (k) = 2 t_{\rm c} \left\{ \cos \( ka \) - 1 \right\} + \frac{D}{2}, \qquad
 \varepsilon_{\rm f} (k) = 2 t_{\rm f} \left\{ \cos \( ka \) - 1 \right\} - \frac{D}{2}, 
 \label{eq:tilde_delta}
\end{gather}
\end{subequations}
and 
\begin{equation}
 \epsilon_0
	= - 2 U_{\rm c} n_{\rm c}^2 - U_{\rm f} \( 1 - 2n_{\rm c} \)^2 - 16 V n_{\rm c} \( 1 - 2n_{\rm c} \)
		+ \frac{2}{V} \sum_{\alpha, \beta} \left| \Delta_{\alpha, \beta} \right|^2 + \frac{1}{2 \lambda} \delta^2,
\end{equation}
where $D$ is the noninteracting band gap when $t_{\rm c}  t_{\rm f} < 0$ and $a$ is the lattice constant.  

Diagonalizing the mean-field Hamiltonian given in Eq.~(\ref{eq:mwjFOxAS}), we obtain
\begin{equation}
 H^{\rm{MF}}
	= \sum_{k, \sigma} \sum_{\epsilon=c, \pm}
		E_{k, \epsilon}
		\gamma^\dagger_{k, \epsilon, \sigma} \gamma_{k, \epsilon, \sigma}
 	+ N \epsilon_0,
\end{equation}
where
\begin{equation}
 E_{k, c} = \tilde{\varepsilon}_{\rm c} (k), \quad
 E_{k, \pm} = \eta_k \pm E_k
\end{equation}
with
\begin{equation}
 \eta_k = \frac{\tilde{\varepsilon}_{\rm c}(k) + \tilde{\varepsilon}_{\rm f}(k) }{2}, \quad
 \xi_k = \frac{\tilde{\varepsilon}_{\rm c}(k) - \tilde{\varepsilon}_{\rm f}(k) }{2}, \quad
 E_k = \sqrt{ \xi_k^2 + \left| \tilde{\Delta}_{1}(k) \right|^2 + \left| \tilde{\Delta}_{2}(k) \right|^2}.
\end{equation}
$\gamma_{k, \epsilon, \sigma}$ ($\gamma^\dagger_{k, \epsilon, \sigma}$) is the annihilation 
(creation) operator of the quasiparticle.
The quasiparticle operators satisfy the relation 
$c_{k, \mu, \sigma} = \sum_{\epsilon} \psi_{k, \sigma; \mu, \epsilon} \gamma_{k, \epsilon, \sigma}$, 
where $\psi_{k, \sigma; \mu, \epsilon}$ is the Bogoliubov transformation coefficient 
and $f_{k, \sigma} = c_{k, 3, \sigma}$.
The coefficients $\psi_{k, \sigma; \mu, \epsilon}$ are determined as 
\begin{equation}
 \begin{pmatrix}
	c_{k, 1, \sigma} \\ c_{k, 2, \sigma} \\ f_{k, \sigma}
 \end{pmatrix}
 =
 \begin{pmatrix}
	\frac{\tilde{\Delta}_{1}^*}{\left| \tilde{\Delta}_{1} \right|} \frac{\left| \tilde{\Delta}_{2} \right|}{\sqrt{\left| \tilde{\Delta}_{1} \right|^2 + \left| \tilde{\Delta}_{2} \right|^2}} &
	- \frac{\tilde{\Delta}_1^*}{2 E_k v_k} &
	\frac{\tilde{\Delta}_1^*}{2 E_k u_k}
	\\
	- \frac{\tilde{\Delta}_{2}^*}{\left| \tilde{\Delta}_{2} \right|} \frac{\left| \tilde{\Delta}_{1} \right|}{\sqrt{\left| \tilde{\Delta}_{1} \right|^2 + \left| \tilde{\Delta}_{2} \right|^2}} &
	- \frac{\tilde{\Delta}_2^*}{2 E_k v_k} &
	\frac{\tilde{\Delta}_2^*}{2 E_k u_k}
	\\
	0 &
	v_k &
	u_k
 \end{pmatrix}
 \begin{pmatrix}
	\gamma_{k, c, \sigma} \\ \gamma_{k, +, \sigma} \\ \gamma_{k, -, \sigma}
 \end{pmatrix}
 \label{eq:Bogoliubov_trans}
\end{equation}
with
\begin{equation}
 u_{k} = \sqrt{\frac{1}{2} \( 1 + \frac{\xi_{k}}{E_{k}} \)}, \quad
 v_{k} = \sqrt{\frac{1}{2} \( 1 - \frac{\xi_{k}}{E_{k}} \)}.
\end{equation}

According to the Ginzburg-Landau theory, the order parameters are determined to minimize the free energy 
\begin{equation}
F = 2 \mu N - k_{\rm B} T \sum_{k, \epsilon, \sigma}
 	\ln \( 1 + \e^{-\beta \( E_{k, \epsilon} - \mu \)} \)
 	+ N \epsilon_0,
 \label{eq:syKEP7pN}
\end{equation}
where $\mu$ is a chemical potential determined from the electron number conservation $2 N = \sum_{k, \sigma} \sum_{\epsilon=c, \pm} f(E_{k, \epsilon})$.  
$f(E) = \frac{1}{\e^{\beta \(E - \mu \)} + 1}$ is the Fermi distribution function at reciprocal temperature $\beta = 1/ k_{\rm B}T$ with the Boltzmann constant $k_{\rm B}$.
If an order parameter is written as $x$, it satisfies the equation $\der F/ \der x = 0$.
This equation reads
\begin{equation}
 \sum_{k, \epsilon, \sigma}
	\frac{\der E_{k, \epsilon}}{\der x} f (E_{k, \epsilon})
 	+ N \frac{\der \epsilon_0}{\der x}
	= 0, 
\end{equation}
from which we obtain the following self-consistent equations:  For the number of conduction electrons $n_{\rm c}$,  
\begin{equation}
   n_{\rm c} = \frac{1}{2N} \Bigl\{\sum_{k} f (E_{k, c})
 	+ \sum_{k} u_k^2 f (E_{k, +})
	+ \sum_{k} v_k^2 f (E_{k, -}) \Bigr\},
\end{equation}
for the excitonic order parameters $\Delta_{\alpha, 1}$ and $\Delta_{\alpha, 2}$,
\begin{align}
 \Delta_{\alpha, 1}
	&=  - \frac{V}{2N} \sum_{k} \frac{\tilde{\Delta}_\alpha}{E_k} \( f(E_{k, +}) - f(E_{k, -}) \) \\
 \Delta_{\alpha, 2}
	&=  - \frac{V}{2N} \sum_{k} \e^{\mp \i k a}
		\frac{\tilde{\Delta}_\alpha}{E_k} \( f(E_{k, +}) - f(E_{k, -}) \),
\end{align}
and for the displacement of the Ta atoms $\delta$,
\begin{equation}
 \delta
	= - \frac{\lambda}{N} \sum_{k} \frac{\tilde{\Delta}_{1} + \tilde{\Delta}_{1}^* + \tilde{\Delta}_{2} + \tilde{\Delta}_{2}^* }{E_k}
 		\( f(E_{k, +}) - f(E_{k, -}) \).
\end{equation}
We solve these equations self-consistently to obtain the order parameters.  

The parameters in our model are the same as those in Ref.~\cite{Kaneko2013PRB}.
We set the hopping integrals as $t_{\rm c} = -0.8$~eV and $t_{\rm f} = 0.4$~eV to reproduce 
the conduction and valence bandwidths obtained from the density-functional-theory-based 
electronic structure calculation \cite{Kaneko2013PRB}.  The noninteracting band gap $D = 0.2$ eV is estimated from experiment.  
We assume the relation $U_{\rm c} = U_{\rm f} = 4V$ in order to minimize the Hartree sift in the mean-field approximation.  

\begin{figure}
\begin{center}
\includegraphics[width=8.0cm]{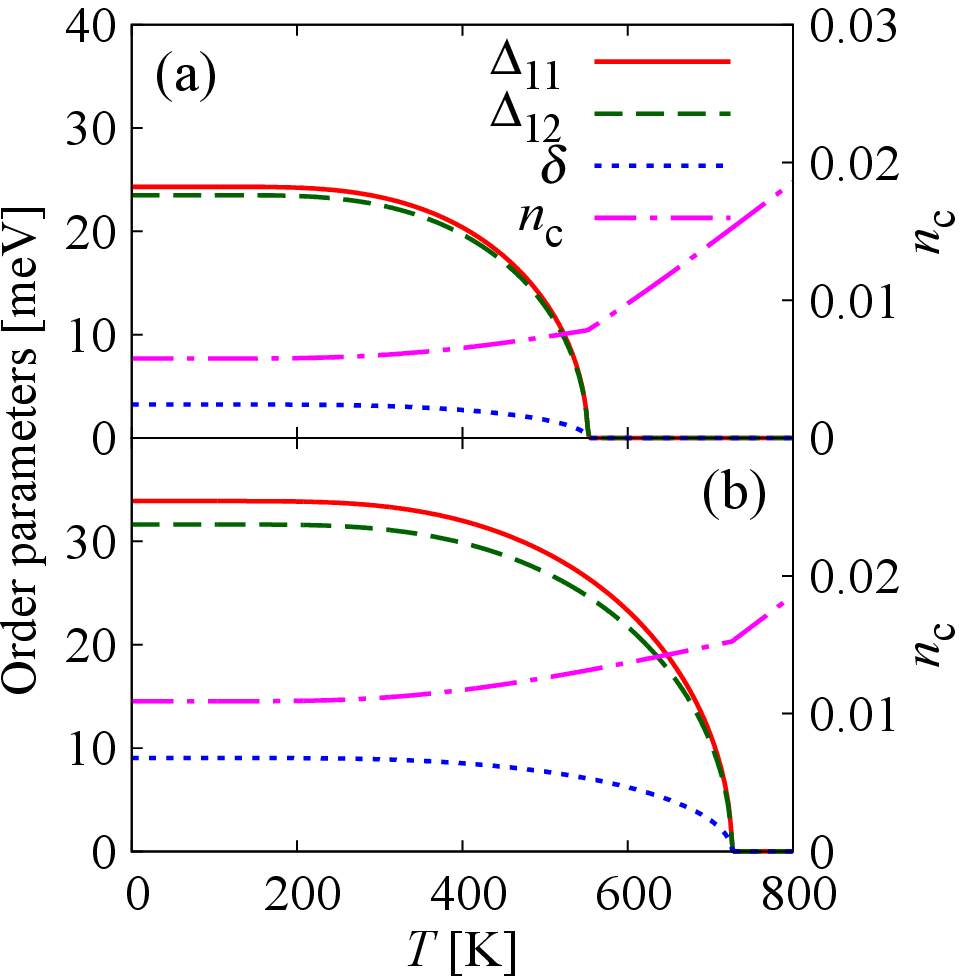}
\caption{(Color online)
Calculated temperature dependence of the order parameters and the number of conduction electrons, 
where we assume $V = 0.6$~eV and $\lambda = 0.01$~eV in (a) and $\lambda = 0.02$~eV in (b).  
The red solid line is for $\Delta_{1,1}$, the green dashed line is for $\Delta_{1,2}$, 
blue dotted line is for $\delta$, and magenta dot-dash line is for $n_{\rm c}$.
Note that $\Delta_{1,1} = \Delta_{2,1}$ and $\Delta_{1,2} = \Delta_{2,2}$ 
because of the equivalence of the two Ta chains $\alpha=1$ and $2$.  
}
\label{fig:op-temp}
\end{center}
\end{figure}

Figure~\ref{fig:op-temp} shows the temperature dependence of the order parameters 
$\Delta_{1,1}$, $\Delta_{1,2}$, and $\delta$ and the number of CB electrons $n_{\rm c}$.  
Note that the excitonic condensation and structural phase transition occurs 
simultaneously because the order parameters enter Eq.~(\ref{eq:tilde_delta}) as a sum 
of the order parameters for the excitonic condensation and the lattice distortion.  
Note also that $n_{\rm c}$ is finite even at zero temperature in the excitonic phase.  
We estimate $T_{\rm c} \simeq 552$~K at $\lambda = 0.01$~eV from Fig.~\ref{fig:op-temp}(a) 
and $T_{\rm c} \simeq 726$~K at $\lambda = 0.02$~eV from Fig.~\ref{fig:op-temp}(b).  
These values are larger than the experimental value $T_{\rm c}=328$~K \cite{DiSalvo1986JLCM} 
because the mean-field approximation ignores the quantum fluctuations completely.  
See Ref.~\cite{Kaneko2013PRB} for further details of the mean-field analysis of our model.  

\vfill
\eject

